\documentclass[useAMS,usenatbib,usegraphicx]{mn2e}
\usepackage{psfig}   
\usepackage{graphicx}
\usepackage{amssymb}
\usepackage{subfigure}
\usepackage{lineno}

\title[Mass segregation in $\rho$~Ophiuchi]{A search for mass segregation of stars and brown dwarfs in $\rho$~Ophiuchi}

\author[R.~J.~Parker, Th.~Maschberger and C.~Alves de Oliveira]{
  Richard J.~Parker$^1$\thanks{E-mail: rparker@phys.ethz.ch},
  Thomas Maschberger$^2$ and Catarina Alves de Oliveira$^3$ \vspace*{0.1cm}\\
   $^1$ Institute for Astronomy, ETH Z{\"u}rich, Wolfgang-Pauli-Strasse 27, 8093, Z{\"u}rich, Switzerland\\
   $^2$ Institut de Plan{\'e}tologie et d'Astrophysique de Grenoble, BP 53, F-38041 Grenoble C{\'e}dex 9, France\\
   $^3$ Herschel Science Centre, European Space Astronomy Centre (ESA), P.O. Box, 78, 28691 Villanueva de la Ca\~{n}ada, Madrid, Spain}

\begin{document}

\date{}
                             
\pagerange{\pageref{firstpage}--\pageref{lastpage}} \pubyear{2012}

\maketitle

\label{firstpage}


\begin{abstract}
We apply two different algorithms to search for mass segregation to a recent observational census of the $\rho$~Ophiuchi star forming region. Firstly, we apply the $\Lambda_{\rm MSR}$ method, 
which compares the minimum spanning tree (MST) of a chosen subset of stars to MSTs of random subsets of stars in the cluster, and determine the mass segregation ratio, $\Lambda_{\rm MSR}$. Secondly, we apply the $m-\Sigma$ method, which calculates the local stellar surface density 
around each star and determines the statistical significance of the average surface density for a chosen mass bin, compared to the average surface density in the whole cluster. 
Using both methods,  we find no indication of mass segregation (normal or inverse) in the spatial distribution of stars and brown dwarfs in $\rho$~Ophiuchi.
Although $\rho$~Ophiuchi suffers from high visual extinction, we show that a significant mass segregation signature would be detectable, albeit slightly diluted, despite dust obscuration of centrally located massive stars.   
\end{abstract}


\section{Introduction}

Most stars form in groups, clusters, and larger associations. In order to understand the star formation process, it is desirable to quantify the spatial distribution of stars in different star forming regions, 
so that a clear picture of the formation and evolution of each region can be drawn. It is possible to measure the amount of substructure in a region \citep[e.g.\,\,by using 
the $\mathcal{Q}$-parameter,][]{Cartwright04} and to quantify the amount of mass segregation (e.g.\,\,the $\Lambda_{\rm MSR}$ method, \citealp{Allison09a}, or the $m-\Sigma$ method, \citealp{Maschberger11}). Additionally, statistical 
methods can be applied to find clusters against a background field \citep[e.g.][]{Gutermuth09,Schmeja11}.

\citet{Allison09a} found that the amount of mass segregation in the ONC could be quantified by comparing the minimum spanning trees (MSTs) of chosen subsets of stars to the MSTs of random sets of stars. If the MST of the most massive stars is shorter than the MSTs of random subsets of cluster stars, then the cluster is mass segregated. The ONC is 
mass segregated \citep[see also][]{Hillenbrand98}, and the same signature was found by \citet{Sana10} in Trumpler~14. 

However, \citet{Parker11b} found that the most massive stars in the Taurus association were `inversely mass segregated', i.e.\,\,anti-clustered with respect to randomly chosen stars. Is mass segregation therefore a dynamical process \citep[as postulated by][]{Allison09b}, rather than a primordial outcome of star formation (in hydrodynamical simulations of star cluster formation, 
primordial mass segregation occurs as part of the competitive accretion process, e.g.\,\,\citealp{Maschberger10}, \citealp{Maschberger11})? To answer this question, we must first search for mass segregation in other young star forming regions, ideally using independent methods.

\begin{figure*}
\begin{center}
\rotatebox{270}{\includegraphics[scale=0.65]{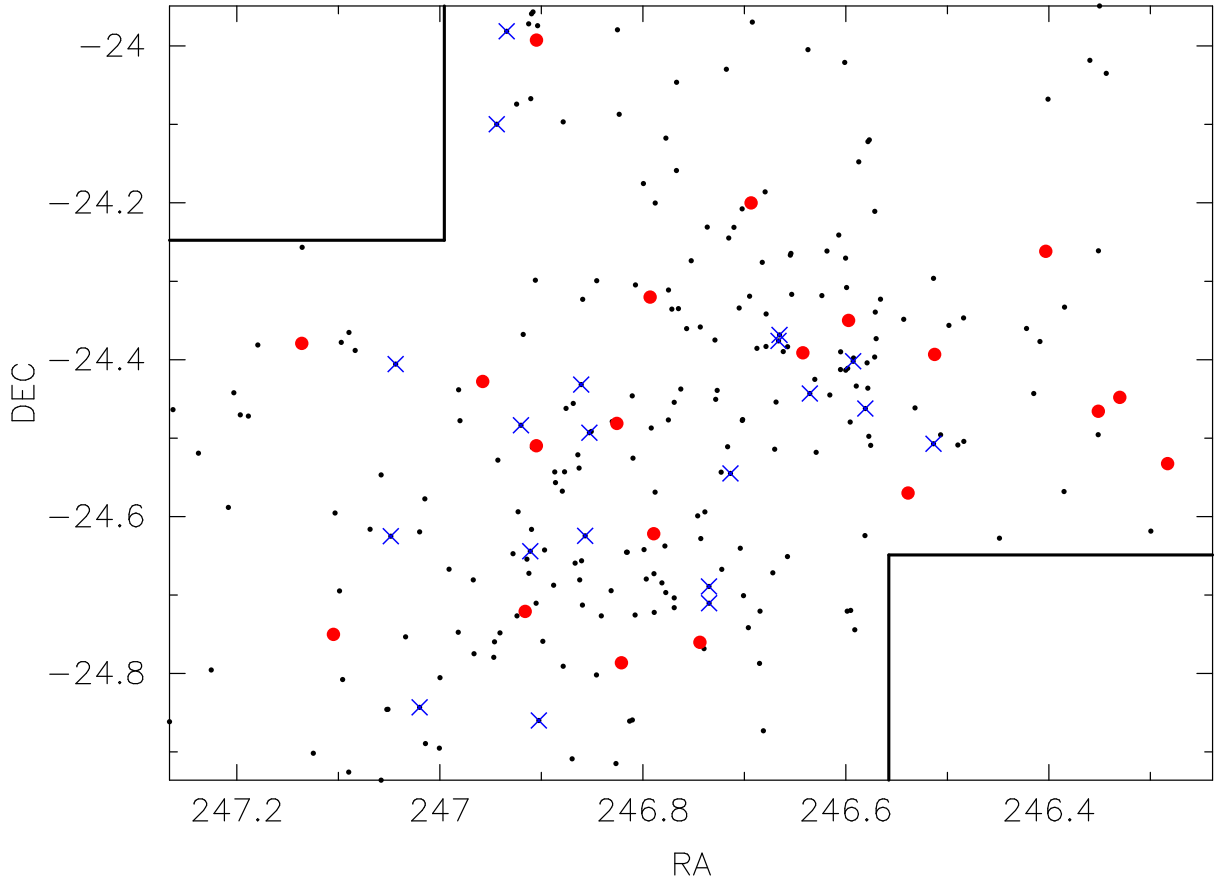}}
\end{center}
\caption[bf]{A map of $\rho$~Ophiuchi showing the 255 objects in our dataset (restricted to the WIRCam field). The 20 least massive cluster 
members (masses up to 0.03\,M$_\odot$) are shown by the blue crosses and the 20 most massive cluster members (masses down to 1.63\,M$_\odot$) are shown by the large red dots. The solid lines indicate the extent of the WIRCam field.}
\label{Oph_map}
\end{figure*}

In this paper we search for mass segregation in $\rho$~Ophiuchi. This cluster suffers heavily from differential extinction, so an accurate and self-consistent determination of stellar masses is difficult. However, recent spectroscopic 
surveys \citep{Alves12,Erickson2011,Geers2011,Muzic2011} have probed the low-mass end of the IMF and allowed a complete census of the cluster to be made. We describe the observational sample in Section~\ref{observations}, we describe the methods used to quantify mass segregation in Section~\ref{method}, we 
present our results in Section~\ref{results}, we discuss the results and the potential effects of extinction in Section~\ref{discuss} and we conclude in Section~\ref{conclude}.   
 
\section{The observational sample}
\label{observations}

We adopted as a starting point for building the observational sample the most recent census of the $\rho$~Ophiuchi core cluster as compiled 
in \citet{Alves12}, where a new population of brown dwarf members in the cluster was uncovered and analyzed 
with respect to the previously known members. Their census includes only stars and brown dwarfs for which there is a spectral type classification 
and reliable membership confirmation. In short, the compilation takes all the spectroscopically confirmed members compiled for the cluster's review 
\citep[see,][ and references therein]{Wilking2008} in the Handbook of Star Forming Regions \citep{Reipurth2008}, adding to it the more recent spectroscopic 
results by \citet{Alvesdeoliveira2010,McClure2010,Geers2011,Erickson2011}, totalling a list of 250 members where 208 have spectral types earlier than M6, 
and 42 have spectral types later or equal to M6. At the age of $\rho$~Oph ($\sim$1~Myr), the evolutionary models of the Lyon group 
\citep{Baraffe1998,Chabrier2000} when combined with the temperature scale of \citet{Luhman2003}, place the substellar boundary at $\sim$M6.25 
\citep{Luhman2007}. We have added to this census two low mass stars and one brown dwarf recently confirmed spectroscopically by \citet{Muzic2011}, as well 
as 3 members where the spectral type carries a larger error (but membership is confirmed) presented by Alves de Oliveira et al. (2012) but not included in their compilation. 

In the substellar regime, the spectroscopic follow-up of the CFHT/WIRCam survey \citep{Alves12} is nearly complete down to an extinction of 20 visual 
magnitudes (only 3 photometric candidates were not observed spectroscopically), and within the WIRCam mapped region (interior to the solid lines in Fig.~\ref{Oph_map}). In the 
stellar domain, a conservative depth of 8 visual magnitudes has been
used by \citet{Alves12} to define a complete sample with spectroscopic confirmation. 
To complete the database in the stellar domain at higher extinctions, we have included candidate members from X-ray surveys that still lack a spectroscopic confirmation. 
X-ray surveys of young stellar objects usually have low contamination rates, in particular in clusters like $\rho$~Oph where large amounts of extinction effectively block 
background sources. From the 51 X-ray sources which lack spectroscopic confirmation compiled in the \citet{Wilking2008} list of candidate members 
\citep[originally uncovered by][]{Imanishi2001,Gagne2004,Ozawa2005,Pillitteri2010}, 43 are matched to a near-IR source, either in the 2MASS or the WIRCam catalogues. 
From the remaining 8 X-ray sources, 2 have an uncertain membership status \citep[denoted as \emph{x?} in][]{Wilking2008}, and 6 are not detected in the \emph{J}-band, 
and in any case are outside the photometric completeness limits of the WIRCam survey \citep[\emph{J}$=$20.5 and \emph{H}$=$18.9~mag,][]{Alves12}. 

We have used the colour-colour diagram \emph{J}$-$\emph{H}~vs.~\emph{H}$-$\emph{K} to deredden each of the X-ray sources along the extinction vector 
\citep{Rieke1985} and estimate a spectral type by comparing their near-IR photospheric colours to those characteristic of young stellar objects 
\citep[their Table~13]{Luhman2010}. This method could not be applied for 9 sources which have strong IR excess and therefore their position on the colour-colour diagram 
is likely to be affected by the contribution of the disk, and to 4 sources which although classified as Class~III fall in a region of the diagram where the colours of young 
stellar objects ($\sim$M9 to early L) increase nearly parallel to the reddening vector, and therefore any solution is degenerate. We estimated spectral types 
($\sim$K4 to $\sim$M8) for 30 sources using this method.

To estimate the masses of the members of the cluster, we first convert spectral types to temperature, adopting the temperature scale from \citet{Schmidt82} for stars 
earlier than M0, and the scale from \citet{Luhman2003} for sources with spectral type between M0 and M9.5. For the L dwarfs, we applied the scale proposed by 
\citet{Lodieu08} extrapolated to the L4 spectral type. Masses were derived from the 1~Myr evolutionary models \citep{Baraffe1998,Chabrier2000,Siess00} according
 to each target's effective temperature.  

Because we derive our masses from temperatures, many stars are assigned the same mass from the stellar models. This is potentially a problem for our mass segregation 
algorithm, as 25 stars at the peak of the IMF may be assigned the same mass. We therefore apply a small amount of random noise to each mass, thereby making each value unique.  

 In Fig.~\ref{Oph_map} we show the 255 objects in our dataset (which we have restricted to the WIRCam field). We plot the 20 most massive stars (masses down to 1.63\,M$_\odot$) 
as the large red points, and the 20 least massive objects (masses up to 0.03\,M$_\odot$) 
as the blue crosses. 

\subsection{Spatially and extinction-limited sample}

A major caveat in studying a representative sample of the $\rho$~Ophiuchi population is the variable extinction across the cluster. To attempt to correct our methods 
for this, we also examined a spatially and extinction limited sample of objects. We selected from the original data base all sources that had in the  colour-magnitude 
diagram \emph{H}~vs.~\emph{J}$-$\emph{H} an A$_{V}$$\lesssim$20~mag \citep[see, for example, Fig.~9 in][]{Alves12}. The data base limited both spatially 
and to an extinction of 20 visual magnitudes contains 205 members. Though the masses of the X-ray members determined from photometry are likely to carry a large 
uncertainty, they represent only $\sim$11$\%$ of this sample, and should nevertheless reflect in relative terms the relation between the real masses. 

\section{Method}
\label{method}

In this section we outline the two methods we use to look for mass segregation signatures in the data, namely the $\Lambda_{\rm MSR}$ 
ratio pioneered by \citet{Allison09a} and the  $m-\Sigma$ distribution, recently proposed by \citet{Maschberger11}.

\subsection{The $\Lambda_{\rm MSR}$ mass segregation ratio}

We first quantify any mass segregation present in the cluster by using the $\Lambda_{\rm MSR}$  ratio introduced by \citet{Allison09a}. This 
constructs a minimum spanning tree (MST) between a chosen subset of stars and then compares this MST to the average MST length of many 
random subsets. 

The MST
of a set of points is the path connecting all the points via the
shortest possible pathlength but which  contains no closed loops
\citep[e.g.][]{Prim57,Cartwright04}.

We use the algorithm of \citet{Prim57} to construct MSTs in our
dataset. We first make an ordered list of the separations  between all
possible pairs of stars\footnote{From this point onwards, when referring in 
general to `stars' in the cluster, we mean `stars \emph{and} brown dwarfs', 
as we are including all the objects in the observational
sample.}. Stars are then connected together in `nodes',
starting with the shortest separations and  proceeding through the
list in order of increasing separation, forming new nodes if the
formation of the node does not result in a closed loop.

We find the MST of the $N_{\rm MST}$ stars in the chosen subset and
compare this to the MST of sets of $N_{\rm MST}$ random  stars in the
cluster. If the length of the MST of the chosen subset is shorter than
the average length of the MSTs for the  random stars then the subset
has a more concentrated distribution and is said to be mass segregated. Conversely, if the MST  length of the chosen subset is
longer than the average MST length, then the subset has a less
concentrated distribution, and is  said to be inversely mass
segregated \citep[see e.g.][]{Parker11b}. Alternatively, if the MST length of the chosen subset is
equal to the random MST length,  we can conclude that no mass
segregation is present.

By taking the ratio of the average (mean) random MST length to the subset MST
length, a quantitative measure of the degree of  mass segregation
(normal or inverse) can be obtained. We first determine the subset MST
length, $l_{\rm subset}$. We then  determine the average length of
sets of $N_{\rm MST}$ random stars each time, $\langle l_{\rm average}
\rangle$. There is a dispersion  associated with the average length of
random MSTs, which is roughly Gaussian and can be quantified as the
standard deviation  of the lengths  $\langle l_{\rm average} \rangle
\pm \sigma_{\rm average}$. However, we conservatively estimate the lower (upper) uncertainty 
as the MST length which lies 1/6 (5/6) of the way through an ordered list of all the random lengths (corresponding to a 66 per cent deviation from 
the median value, $\langle l_{\rm average} \rangle$). This determination 
prevents a single outlying object from heavily influencing the uncertainty. 
We can now define the `mass  segregation ratio' 
($\Lambda_{\rm MSR}$) as the ratio between the average random MST pathlength 
and that of a chosen subset, or mass range of objects:
\begin{equation}
\Lambda_{\rm MSR} = {\frac{\langle l_{\rm average} \rangle}{l_{\rm subset}}} ^{+ {\sigma_{\rm 5/6}}/{l_{\rm subset}}}_{- {\sigma_{\rm 1/6}}/{l_{\rm subset}}}.
\end{equation}
A $\Lambda_{\rm MSR}$ of $\sim$ 1 shows that the stars in the chosen
subset are distributed in the same way as all the other  stars,
whereas $\Lambda_{\rm MSR} > 1$ indicates mass segregation and
$\Lambda_{\rm MSR} < 1$ indicates inverse mass segregation,
i.e.\,\,the chosen subset is more sparsely distributed than the other stars.

As noted by \citet{Allison09a}, the MST method gives a quantitative
measure of mass segregation with an associated significance and   it
does not rely on defining the centre of a cluster.

There are several subtle variations of $\Lambda_{\rm MSR}$. \citet*{Olczak11} propose using the geometric mean to reduce the spread in uncertainties, 
and \citet{Maschberger11} propose using the median MST length to reduce the effects of outliers from influencing the results. However, in the subsequent 
analysis we will adopt the original $\Lambda_{\rm MSR}$  from Allison. 

\subsection{The $m-\Sigma$ distribution}

Recently, \citet{Maschberger11} proposed a method to analyse mass segregation which measures the distribution of local stellar surface density, $\Sigma$, as a function of stellar mass.
We calculate the local stellar surface density following the prescription of \citet{Casertano85}, modified to account for the analysis in projection. For an individual star the local stellar surface density is given by
\begin{equation}
\Sigma = \frac{N - 1} {\pi r_{N}^2},
\end{equation}
where $r_{N}$ is the distance to the $N^{\rm th}$ nearest neighbouring star (we adopt $N = 10$ throughout this work).

If there is mass segregation, massive stars are concentrated in the central, dense region of a cluster and thus should have higher values of $\Sigma$.
This can be seen in a plot of $\Sigma$ versus mass, showing all stars and highlighting outliers.
Trends in the $m-\Sigma$ plot can be shown by the moving average (or median) of a subset, $\tilde{\Sigma}_\mathrm{subset} $, compared to the average (median) of the whole sample, $\tilde{\Sigma}_\mathrm{all}$.
The signature of mass segregation is then  $\tilde{\Sigma}_\mathrm{subset} >  \tilde{\Sigma}_\mathrm{all}$, and of inverse mass segregation  $\tilde{\Sigma}_\mathrm{subset} <  \tilde{\Sigma}_\mathrm{all}$.
The statistical significance of mass segregation can be established with a two-sample Kolmogorov-Smirnov test of the $\Sigma$ values of the subset against the $\Sigma$ values of the rest. \\

Note that there are many more ways of defining mass segregation. For instance, one can choose a cluster centre and measure the mass 
function as a function of radial distance \citep{Gouliermis04,Sabbi08}, or the distance of the most massive star(s) from the cluster centre 
compared to the average distance of low-mass stars to the cluster centre \citep{Kirk10}. Both methods rely on determining the centre of the cluster 
or association, which in the case of low-number clusters with substructure is non-trivial and is virtually impossible in the case of a highly substructured region such as 
Taurus \citep{Parker11b}. 

\section{Results}
\label{results}

In this section we present the results of our $\Lambda_{\rm MSR}$ analysis, followed by the $m-\Sigma$ distribution. We then discuss the effects of extinction on the results. 

\subsection{$\Lambda_{\rm MSR}$  for high mass stars}

\begin{figure}
\begin{center}
\rotatebox{270}{\includegraphics[scale=0.33]{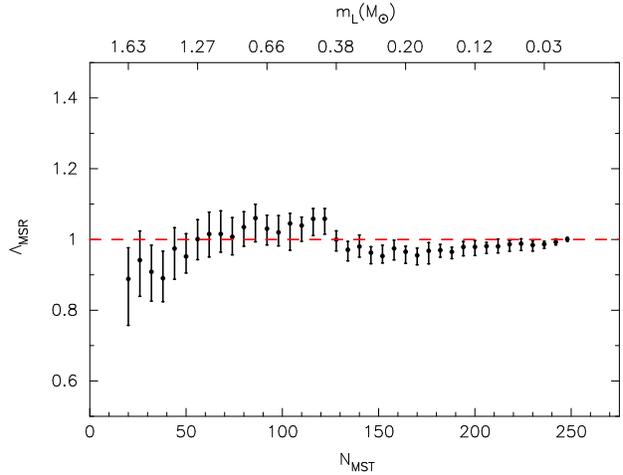}}
\end{center}
\caption[bf]{The evolution of the mass segregation ratio, $\Lambda_{\rm MSR}$, for the $N_{\rm MST}$  most massive stars in our dataset and we also indicate the 
lowest mass star, $m_{\rm L}$ within the $N_{\rm MST}$.  Error bars show the 1/6 and 5/6 percentile values from the median, as described in the text. The dashed 
line indicates $\Lambda_{\rm MSR} = 1$,  i.e.\,\,no mass segregation.}
\label{rho_Oph_hm_mst}
\end{figure}

In Fig.~\ref{rho_Oph_hm_mst} we show the evolution of $\Lambda_{\rm MSR}$ as a function of the number of stars in an MST, $N_{\rm MST}$ for the most massive stars in 
the cluster. We increase the number of stars in the MST in steps of 6, which is a compromise between a high enough resolution to pick out structure between different 
mass regimes, and a low enough resolution so that we do not add noise to the plot. The first subset compares the MST of the 20 most massive stars to the median of 
many different random sets of 20 stars, and the second subset is the 26 most massive stars compared to the median of random sets of 26 stars, and so on. On the top 
axis we also indicate the mass of the least massive star within that value of $N_{\rm MST}$, at regular intervals. 

Fig.~\ref{rho_Oph_hm_mst} we see that there is no clear mass segregation signature (normal or inverse) in the most massive stars in the cluster (the most massive 20 stars are indicated by the large red points in Fig.~\ref{Oph_map}). 
The 20 most massive stars (with masses above 1.63\,M$_\odot$) have a mass segregation ratio $\Lambda_{\rm MSR} = 0.89^{+0.09}_{-0.13}$, which does deviate from $\Lambda_{\rm MSR} = 1$ (indicating slight inverse mass segregation), but because the 26 most massive stars are 
consistent with $\Lambda_{\rm MSR} = 1$, this result is not particularly significant. 

\subsection{$\Lambda_{\rm MSR}$ for low mass stars}

\begin{figure}
\begin{center}
\rotatebox{270}{\includegraphics[scale=0.33]{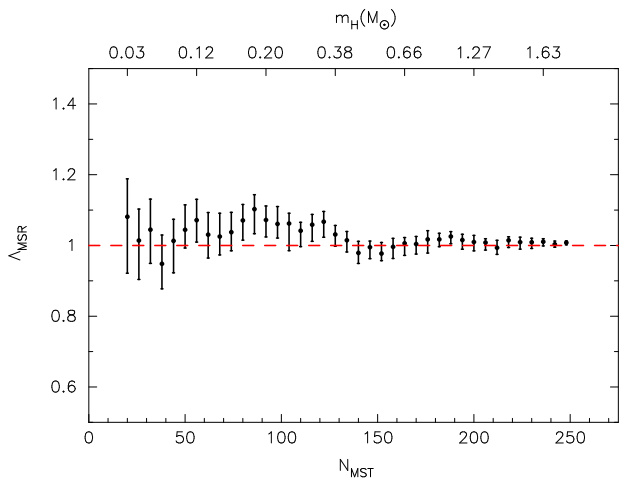}}
\end{center}
\caption[bf]{The evolution of the mass segregation ratio, $\Lambda_{\rm MSR}$, for the $N_{\rm MST}$  least massive stars in our dataset and we also indicate the 
highest mass star, $m_{\rm H}$ within the $N_{\rm MST}$.  Error bars show the 1/6 and 5/6 percentile values from the median, as described in the text. The dashed 
line indicates $\Lambda_{\rm MSR} = 1$,  i.e.\,\,no mass segregation.}
\label{rho_Oph_lm_mst}
\end{figure}

In Fig.~\ref{rho_Oph_lm_mst} we show the evolution of $\Lambda_{\rm MSR}$ as a function of the number of stars in an MST, $N_{\rm MST}$ for the least massive stars 
in the cluster. We begin by constructing an MST  with the 20 least massive objects in the cluster, and then increasing the number of objects in the MST by 6 at each 
stage. On the top axis we now indicate the mass of the most massive star within the $N_{\rm MST}$ subset. 

We see that the least massive objects do not show any strong mass segregation signature, and (within the uncertainties) are consistent with $\Lambda_{\rm MSR} = 1$.

\subsection{The $m-\Sigma$ distribution}

\begin{figure}
\begin{center}
\includegraphics[scale=0.9]{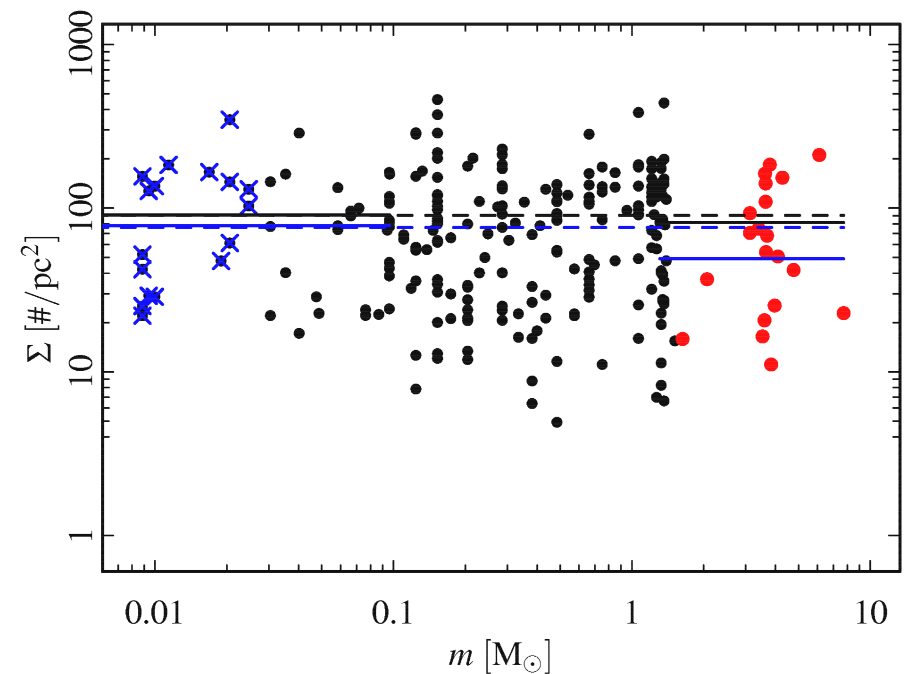}
\end{center}
\caption[bf]{The $m-\Sigma$ distribution for the stars in our dataset. We plot the local surface density for each star against its mass. We then determine the 
median (blue lines) and mean (black lines) $\Sigma$ for the entire cluster (the dashed lines) and for the 50 least massive, and 50 most massive stars in the cluster (the solid lines).}
\label{rho_Oph_msigma}
\end{figure}

We show the $m-\Sigma$ distribution for the stars in our dataset in Fig.~\ref{rho_Oph_msigma}. The upper (black) dashed line is the mean $\Sigma$ value for the whole cluster, and the lower (blue) dashed line is the median value. We also show the mean and median $\Sigma$ values for the 50 most massive stars (on the righthand side) and the 50 least massive stars (on the lefthand side) by the solid lines.

The plot shows that the mean and median $\Sigma$  values of the lowest mass objects in the cluster are marginally higher than for the whole sample. The $p$-values of a two-sample KS test ($\Sigma$ of low-mass stars versus the entire cluster) are $p=0.21$ (20 least massive) and  $p=0.51$ (50 least massive). Usually, these would need to be smaller than $p=0.05$ at a significance level corresponding to $2\sigma$, in order to reject the hypothesis of ``no mass segregation''. Thus, the lowest mass objects are not mass segregated.

The most massive stars lie at slightly lower $\Sigma$ values compared to the whole cluster, suggesting inverse mass segregation. Here the $p$-values are $p=0.17$ and $p=0.70$ for the 50 and 20 most massive stars, respectively. Again, this does not indicate any significant deviation of the spatial distribution of the massive stars from the spatial distribution of the other stars. The 50 most massive stars are inversely mass segregated, similar to the $\Lambda_{\rm MSR}$ results, but only at a $1\sigma$ level.
This is not the case for the 20 most massive stars, where no inverse mass segregation can be concluded.
Given the small $n$ and the rather weak signature for $\Lambda_{\rm MSR}$ this result can be deemed compatible with the $\Lambda_{\rm MSR}$.

\subsection{Extinction-limited sample}

A major caveat in determining the spatial distribution of a sample of objects in $\rho$~Ophiuchi is the variable extinction across the cluster. As a check that our results do not change 
when an extinction limit is imposed on the data, we apply an $A_v$ limit of 20 mag and then repeat the MST and $m-\Sigma$ analysis on this extinction-limited sample. We find 
no discernible difference to the results in either case; i.e.\,\,there is no clear mass segregation signature in either the high- or low-mass objects in the cluster.

\begin{figure*}
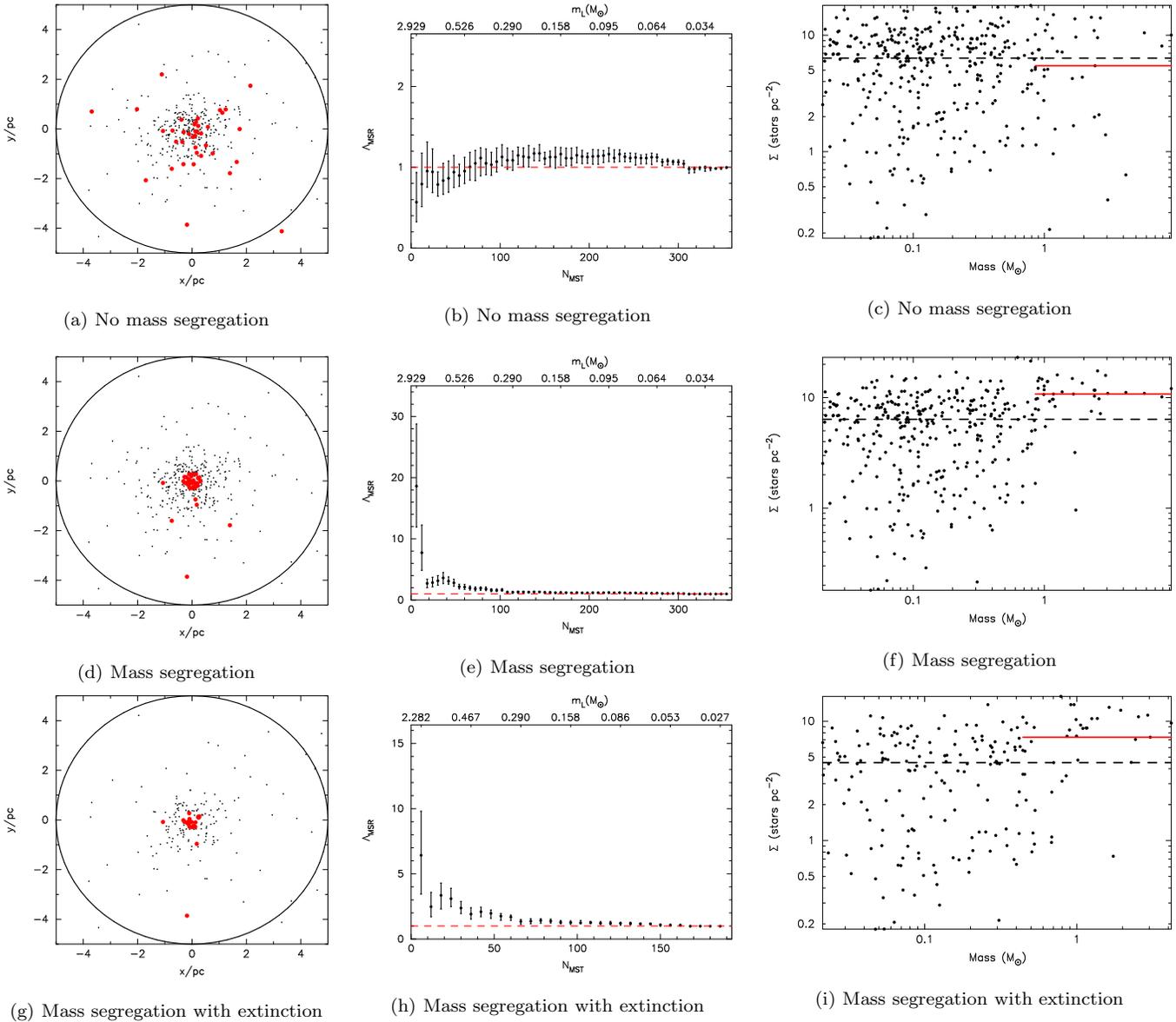

  \begin{center}
\setlength{\subfigcapskip}{10pt}
\hspace*{-0.5cm}\subfigure[No mass segregation]{\label{extinct-a}\rotatebox{270}{\includegraphics[scale=0.24]{fake_map_no_seg.ps}}}
\hspace*{0.3cm} 
\subfigure[No mass segregation]{\label{extinct-b}\rotatebox{270}{\includegraphics[scale=0.23]{fake_mst_no_seg.ps}}}
\hspace*{0.3cm} 
\subfigure[No mass segregation]{\label{extinct-c}\rotatebox{270}{\includegraphics[scale=0.25]{fake_mSig_no_seg.ps}}}
\vspace*{-0.2cm}
\hspace*{-0.5cm}\subfigure[Mass segregation]{\label{extinct-d}\rotatebox{270}{\includegraphics[scale=0.24]{fake_map_seg.ps}}}
\hspace*{0.3cm} 
\subfigure[Mass segregation]{\label{extinct-e}\rotatebox{270}{\includegraphics[scale=0.23]{fake_mst_seg.ps}}}
\hspace*{0.3cm} 
\subfigure[Mass segregation]{\label{extinct-f}\rotatebox{270}{\includegraphics[scale=0.25]{fake_mSig_seg.ps}}}
\vspace*{-0.2cm}
\hspace*{-0.5cm}\subfigure[Mass segregation with extinction]{\label{extinct-g}\rotatebox{270}{\includegraphics[scale=0.24]{fake_map_seg_ex.ps}}}
\hspace*{0.3cm} 
\subfigure[Mass segregation with extinction]{\label{extinct-h}\rotatebox{270}{\includegraphics[scale=0.23]{fake_mst_seg_ex.ps}}}
\hspace*{0.3cm} 
\subfigure[Mass segregation with extinction]{\label{extinct-i}\rotatebox{270}{\includegraphics[scale=0.25]{fake_mSig_seg_ex.ps}}}
\caption[bf]{Results for a fake cluster without mass segregation (panels a -- c), with mass segregation (d -- f), with mass segregation but some stars removed due to high extinction (g -- i). In 
each case we show the morphology of the cluster (the most massive stars are shown by the large red dots), the plot of $\Lambda_{\rm MSR}$ for the high mass stars, and the $m-\Sigma$ plot. The 
extent of the extinction cloud is shown by the black circle. in the $m-\Sigma$ plots the median $\tilde{\Sigma}$ value for the most massive stars is shown by the solid red line, and the median 
$\tilde{\Sigma}$ value for the whole cluster is shown by the dashed line.}
\label{fake_extinct}
  \end{center}
\end{figure*}

\section{Discussion}
\label{discuss}

The results presented in Section~\ref{results} show that there is no evidence of mass segregation in that the most massive stars are not centrally concentrated, as they are in e.g.~the ONC 
\citep{Allison09a} and Trumpler~14 \citep{Sana10}. This could indicate that mass segregation may be a dynamical process, rather than a primordial outcome of star formation, but a study of more star forming regions 
is required to substantiate this hypothesis. 

In this dynamical scenario, the massive stars form at random locations in a substructured cluster, and then a subvirial collapse facilitates mass segregation on a very short timescale \citep[$\sim$ 1\,Myr,][]{Allison09b}. 
$\rho$~Oph is not substructured \citep{Cartwright04}, but may have been at earlier ages. If it was substructured at earlier ages, this has not facilitated dynamical mass segregation in this cluster. 

\subsection{Extinction}

The high level of extinction makes observing objects in $\rho$~Oph challenging, and it is possible that even with our extinction-limited sample, some stars are still hidden in the centre of the cluster. In such a scenario, 
unobserved high mass stars could reside in the central regions, and any mass segregation of such stars would not be observed. In this case, both our mass segregation-finding  algorithms would erroneously give a null-result, 
similar to those described in the previous Section.

Here, we conduct a simple numerical experiment to determine how much a mass segregation signature could be diluted by high levels of extinction, such as that present in $\rho$~Oph. We distribute 360 stars randomly in a Plummer sphere \citep{Plummer11}, 
with a half-number radius of 1\,pc according to the prescription in \citet*{Aarseth74}, and assign masses (again at random) from a 3-part \citet{Kroupa02} IMF of the form:
\begin{equation}
 N(M)   \propto  \left\{ \begin{array}{ll} 
 M^{-0.3} \hspace{0.4cm} m_0 < M/{\rm M_\odot} \leq m_1   \,, \\ 
 M^{-1.3} \hspace{0.4cm} m_1 < M/{\rm M_\odot} \leq m_2   \,, \\ 
 M^{-2.3} \hspace{0.4cm} m_2 < M/{\rm M_\odot} \leq m_3   \,,
\end{array} \right.
\end{equation}
and we choose $m_0$ = 0.02\,M$_\odot$, $m_1$ = 0.1\,M$_\odot$, $m_2$ = 0.5\,M$_\odot$, and  $m_3$ = 10\,M$_\odot$. In Fig.~\ref{extinct-a} we show the morphology of this cluster, with the 40 most massive stars shown by the red points. If we determine $\Lambda_{\rm MSR}$ 
for the most massive stars (in steps of 6 objects) we see that this cluster is not mass segregated, with $\Lambda_{\rm MSR} \simeq 1$ throughout (Fig.~\ref{extinct-b}). The $m-\Sigma$ algorithm also shows no significant differences between the 40 most massive stars and the 
cluster as a whole (the solid red line and the dashed line, respectively, shown in Fig.~\ref{extinct-c}).

We apply a simple mass segregation algorithm to the Plummer sphere by swapping the positions of the 40 most central stars with the positions of the 40 most massive stars (we choose 40 stars to clearly demonstrate the effects of extinction in Fig.~\ref{fake_extinct}, but the results are similar for the 20 most massive stars). 
We show the new spatial configuration of the massive stars in the cluster in Fig.~\ref{extinct-d}. Several of the most massive stars are originally within the sample of the 40 most central stars, and end up (randomly) being assigned positions outside the central core. In one sense, such a configuration is perhaps more 
realistic than if the 40 most massive stars were also the 40 most central; in a real cluster dynamical interactions between the central stars would likely eject one or two of the massive stars. 

In Fig.~\ref{extinct-e} we show the evolution of $\Lambda_{\rm MSR}$ as a function of the number of stars in the MST. The effect of artificially mass segregating the cluster is clearly seen, with $\Lambda_{\rm MSR} = 18.6^{+10.2}_{-6.7}$. The cluster shows significant mass segregation down to the $40^{\rm th}$ most massive star, which 
has $\Lambda_{\rm MSR} = 3.6^{+1.3}_{-0.9}$. Similarly, the $m-\Sigma$ method also shows that the cluster is mass segregated; in Fig.~\ref{extinct-f} we show the median surface density of the entire cluster by the dashed line ($\tilde{\Sigma} = 6.35$~stars\,pc$^{-2}$) and the median surface density of the 40 most massive stars by the red solid line 
($\tilde{\Sigma} = 10.76$~stars\,pc$^{-2}$). A two-sample KS test returns a p-value of $< 10^{-8}$ that the two distributions could be drawn from the same parent population.

We now assign a power-law extinction to the fake cluster, from the centre out to a radius of 5\,pc (denoted by the circle in Figs~\ref{extinct-a},~\ref{extinct-d}~and~\ref{extinct-g}). We then assign an $A_v$ value to each star using the following formula:
\begin{equation}
A_v(r) = 20\left[1 - \left(\frac{|r|}{5\,{\rm pc}}\right)^{5/3}\right],
\end{equation}
where $r$ is the position of the star with respect to the cluster centre and $|r|$ is the modulus of its vector. To account for projection effects along the line of sight, we double $A_v(r)$ if 
the $z$-component of the vector $r$ is negative. Therefore, in the central regions of the cluster, the $A_v$ value can range between $\sim 10 - 40$. We then remove all stars with $A_v > 20$, leaving a total of 193 stars. In Fig.~\ref{extinct-g} we show the spatial distribution of the remaining objects. 

Once again, we calculate $\Lambda_{\rm MSR}$ for the remaining objects, and Fig.~\ref{extinct-h} shows that the mass segregation signature is still observable, although to a lesser extent due to the removal of several of the most massive stars in the cluster. The peak value is now $\Lambda_{\rm MSR} = 6.4^{+3.4}_{-3.0}$, but the 
plot still shows the same morphology as the non-extinction-limited data-sample. Furthermore, the $m-\Sigma$ method also shows that the cluster is still mass segregated; in Fig.~\ref{extinct-i} we show the median surface density of the entire cluster by the dashed line 
($\tilde{\Sigma} = 4.50$~stars\,pc$^{-2}$) and the median surface density of the 40 most massive stars by the red solid line ($\tilde{\Sigma} = 7.34$~stars\,pc$^{-2}$). A two-sample KS test returns a p-value of $< 10^{-2}$ that the two distributions could be drawn from the same parent population. 

We have demonstrated with a simple model for extinction that the two mass segregation finding algorithms could still determine whether a cluster suffering from extinction is significantly mass segregated or not. The results suggest that the actual dataset, whilst possibly lacking some cluster members due to obscuration, is likely reflecting the 
true spatial distribution of stars and brown dwarfs in $\rho$~Oph.

\section{Conclusions}
\label{conclude}

We have used an observational census of $\rho$~Ophiuchi, which was recently enhanced by several surveys probing the substellar domain of the IMF, to search for possible mass segregation 
signatures in the spatial distribution of stars and brown dwarfs in this cluster. 

We have utilised two different algorithms. Firstly, we used the $\Lambda_{\rm MSR}$ technique \citep{Allison09a}, which compares 
the minimum spanning tree (MST) of a chosen subset of stars, to the MSTs of randomly chosen stars in the cluster. If the MST length of a chosen subset is shorter than the MST length of the random objects, 
then the cluster is mass segregated.  Secondly, we have used the $m-\Sigma$ plot, which compares the local 
surface density surrounding massive stars to the the average surface density of all of the stars in the cluster. By this definition, a cluster is mass segregated if the massive stars have a significantly higher than average surface density. Our conclusions are as follows:

(i) The $\Lambda_{\rm MSR}$ technique finds that the most massive stars show hints of being inversely mass segregated, with $\Lambda_{\rm MSR} = 0.89^{+0.09}_{-0.13}$ for the 20 most massive stars. However, $\Lambda_{\rm MSR}$ is consistent with there being no mass segregation 
of the 26 most massive stars, and so on.  The least massive stars show no clear deviation from $\Lambda_{\rm MSR} = 1$. \newline
(ii) The $m-\Sigma$ distribution also suggests that the most massive stars may be inversely mass segregated (but with no strong statistical significance), and with no difference in the distribution of low-mass stars compared to the cluster average. \newline
(iii) The high levels of extinction in $\rho$~Oph may mean that some members are missing from the dataset. However, we have demonstrated that a significant difference in the spatial distribution of a group of objects would still be found by both the  $\Lambda_{\rm MSR}$ and $m-\Sigma$ methods.\newline

In order to understand the star formation process in different clusters, we suggest applying both mass segregation algorithms in tandem to build up a census of the spatial distribution of stars in different star forming regions.

\section*{Acknowledgements}

We thank the referee, Simon Portegies Zwart, for a helpful review. We also thank Jerome Bouvier and Estelle Moraux for their feedback on an earlier draft of this work, 
and Michael Meyer and Vincent Geers for general discussions regarding $\rho$~Oph.

\bibliographystyle{mn2e}
\bibliography{rho_oph_ref}

\label{lastpage}

\end{document}